\documentclass[final,3p,times,twocolumn]{elsarticle}
\usepackage{amsmath}
\usepackage{graphics}
\usepackage{graphicx}
\usepackage{amssymb}
\usepackage{amsthm}
\usepackage{lineno}
\usepackage{placeins}
\usepackage{float}
\usepackage[latin1]{inputenc}

\journal{Nuclear Instruments and Methods in
Physics Research A}
\begin{document}

\begin{frontmatter}



\title{Calibration of the ISOLDE acceleration voltage using a high-precision voltage divider and applying collinear fast beam laser spectroscopy}
\author{A.~Krieger\corref{cor1}\fnref{label1}}
\ead{kriegea@uni-mainz.de}
\cortext[cor1]{Corresponding author at Institut für Kernchemie, Johannes Gutenberg - Universität Mainz, Fritz-Straßmann-Weg 2, 55128 Mainz, Germany.}
\author[label1,label2]{Ch.~Geppert\corref{0}}
\author[label3]{R.~Catherall}
\author[label4]{F.~Hochschulz}
\author[label1]{J.~Krämer}
\author[label1]{R.~Neugart}
\author[label4]{S.~Rosendahl}
\author[label3]{J.~Schipper}
\author[label3]{E.~Siesling}
\author[label4]{Ch.~Weinheimer}
\author[label5]{D. T. ~Yordanov}
\author[label1,label2]{and W.~Nörtershäuser}
\address[label1]{Institut f\"ur Kernchemie, Universit\"at Mainz, 55128 Mainz, Germany}
\address[label2]{GSI Helmholtzzentrum f\"ur Schwerionenforschung, 64291 Darmstadt, Germany}
\address[label3]{CERN, CH-1211 Gen\'{e}ve 23, Switzerland}
\address[label4]{Institut für Kernphysik, Universit\"at M\"unster, 48149 M\"unster, Germany}
\address[label5]{Max-Planck-Institut für Kernphysik, 69117 Heidelberg, Germany}
\begin{abstract}
A high-voltage divider with accuracy at the ppm level and collinear laser
spectroscopy were used to calibrate the high-voltage installation at the
radioactive ion beam facility ISOLDE at CERN. The accurate knowledge of this
voltage is particularly important for collinear laser spectroscopy
measurements. Beam velocity measurements using frequency-comb based collinear laser spectroscopy
agree with the new calibration. Applying this, one obtains consistent results for isotope shifts of stable magnesium isotopes 
measured using collinear spectroscopy and laser spectroscopy on laser-cooled ions in a 
trap. The long-term stability and the
transient behavior during recovery from a voltage dropout were investigated
for the different power supplies currently applied at ISOLDE.
\end{abstract}
\end{frontmatter}

\section{Introduction}

\label{sec:introduction} The accurate determination of high voltages is
crucial for a variety of atomic, nuclear and particle physics experiments. Examples are the
acceleration voltage of ion beams for collinear laser spectroscopy \cite{Kau76}, the
energy of electrons in an electron cooler applied at storage rings to narrow
the momentum distribution of the circulating ion beam \cite{Bud76,Pot90}, or
the voltage applied in a retardation spectrometer to determine the kinetic
energy of electrons emitted in $\beta -$decay \cite{Ott08,Kas04}. Fast-beam experiments
often require precise knowledge of the kinetic energy or the velocity of the
beam particles, especially if resonance processes are investigated. On the
other hand, such resonance processes can be used to measure the ion beam
velocity. In this case, sharp resonances are preferred since the linewidth
is one of the limiting factors for the accuracy that can be achieved.
Collinear laser spectroscopy (CLS) on fast ion or atomic beams is a
particularly clear example for both cases. In the past, it has been
used extensively for studying nuclear ground state properties of short-lived
radioactive isotopes \cite{Neu85,Neu87,Ott89,Bil95,Klu03}. CLS utilizes the
Doppler shift of the frequency of a light wave in the moving frame of an
accelerated ion. Hence, a fixed-frequency laser can be brought into
resonance and tuned across the transition frequency of the ion by
changing the ion velocity. To extract information on isotope shifts and
hyperfine splittings, accurate knowledge of this velocity, {\it{i.e.}} of the ion beam energy, is required.
Typically, voltages of 30-60~kV are applied to the ion source. For most
investigations of medium-mass and heavy nuclei, it is sufficient to know the
voltage applied to the ion source and the potentials at secondary acceleration
or deceleration stages with a relative accuracy of the order $10^{-4}$.
This is the level for which high-voltage dividers are commercially
available. However, as it will be discussed in the first section of this
paper, this accuracy is by far not sufficient for measurements on
the very light elements reaching from hydrogen to roughly neon.

If the frequency of an atomic resonance is well known, CLS can be
used (in reverse) to determine the ion velocity, hence, the kinetic energy and the
voltage applied to the source. This has already been suggested by Poulsen in
1982 \cite{Pou82,Pou88}. The latest development along these lines was a
test apparatus for measurements of high voltages up to a few 100~kV, to
determine and calibrate the voltage of the electron cooler at the GSI
experimental storage ring ESR \cite{Goe04}. The layout of the
prototype allowed measurements up to 50~kV and the $3d_{3/2,5/2}\rightarrow
4p_{3/2}$ transitions in calcium ions were used for probing the ion velocity.
It was demonstrated that accuracies in the order of $10^{-4}$ can be reached
and suggestions for further improvement were made.\\
In this paper, we report on the calibration of the high-voltage dividers
used at ISOLDE/CERN to measure the potential applied to the
high-voltage platforms on which the ion sources are mounted. 
There are two independent units, HT1 and HT2, that can alternatively be used to provide and to measure acceleration voltages up to 60 kV.
The calibration
was performed using a non-commercial high-precision high-voltage divider with accuracy in
the ppm range \cite{Tue07}, and collinear laser spectroscopy on beryllium ions. Both
approaches gave consistent results and the calibration resolved
problems in the earlier analysis of isotope shift measurements on stable and radioactive magnesium
isotopes. 
These problems could be traced back to differences in the measured voltages of HT1 and HT2, which in CLS translate into different Doppler shifts of the resonance frequency. We will start with a short summary of the principles of CLS and their implications for isotope shift measurements, and we will describe test measurements of absolute Doppler shifts to determine the ion velocity.
Then, a concise description of the high-accuracy high-voltage divider will be given and the
results of two calibration campaigns are discussed in results of the long-term as
well as the transient behavior of the high-voltage supplies.

\section{Collinear Laser Spectroscopy}\label{collsec}

CLS was proposed by Kaufman \cite{Kau76} and was rapidly
developed into a useful tool, namely for the spectroscopy of short-lived
radioactive isotopes. Its first on-line installation was realized at a
nuclear research reactor \cite{Sch78} and soon thereafter the COLLAPS (collinear laser spectroscopy)
beamline at the radioactive ion beam facility ISOLDE was
built \cite{Mue83}. Followed by similar and more advanced experiments (see, {\it{e.g.}} \cite{Ulm85,Sil88,Bor89,Gei00,Cam02}) collinear laser
spectroscopy developed into a general technique for on-line investigations
of spins, electromagnetic moments and nuclear charge radii of short-lived
isotopes. It was continuously improved
and combined with different detection techniques and, thus, became
increasingly sensitive and accurate. The latest steps in these
developments where the combination of CLS with an RFQ cooler and buncher
\cite{Nie01} and frequency-comb-based collinear spectroscopy \cite{Noe09,Zak10}.

\begin{figure*}
\centering
\includegraphics[width=\linewidth,height=0.5\linewidth ]{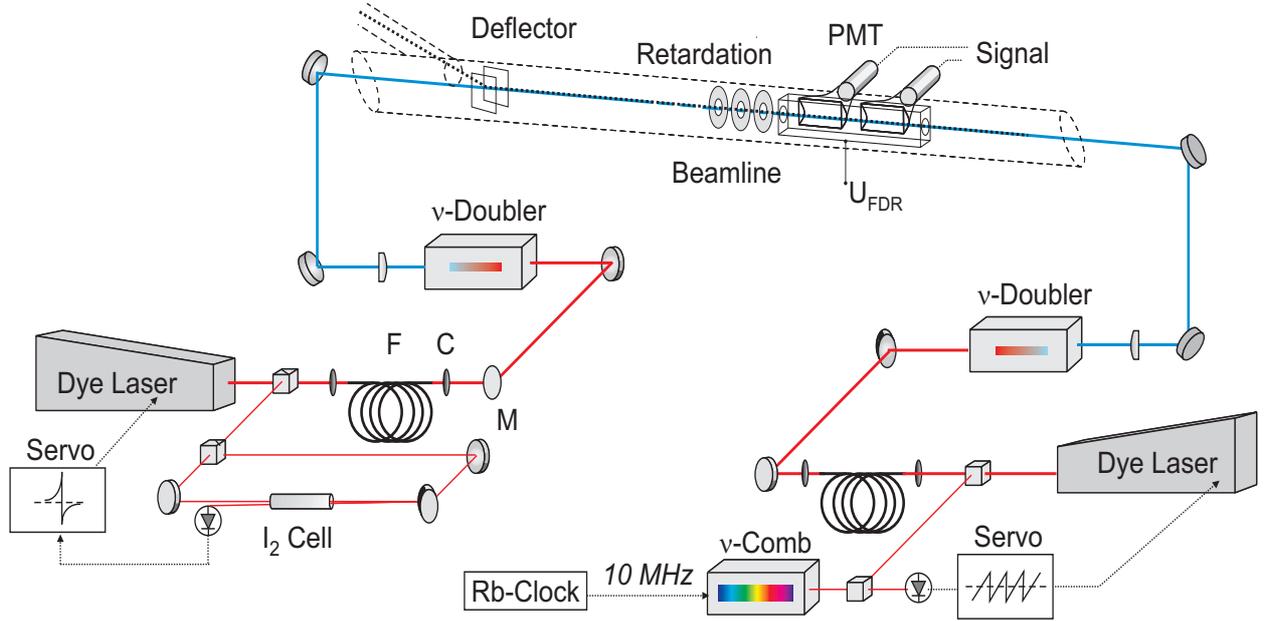}
\caption{Experimental setup for collinear and anticollinear spectroscopy on fast Be$^{+}$-ions at ISOLDE. Two fiber-coupled
dye lasers seed second-harmonic generator rings producing UV light at wavelengths of 312~nm and 314~nm respectively. The laser beams are
superimposed collinearly and anticollinearly with the ion beam. Scanning over a resonance is achieved by applying a small stepwise voltage $U_{FDR}$ to the fluorescence detection region FDR.
region FDR }
\label{collsetup}
\end{figure*}

The scheme of our setup for CLS on an ion beam is shown in
Fig.~\ref{collsetup}. An ion beam at a usual beam energy of
30-60~keV is superimposed with a laser beam in the collinear or anticollinear
direction using an electrostatic deflector. In the figure the special layout
for simultaneous collinear and anticollinear measurements with two laser systems is depicted,
whereas in standard CLS only a single system is applied. Photons are collected in the fluorescence detection region (FDR) and detected by photomultiplier tubes or other
devices capable of single-photon counting. Usually, the ions are
accelerated or decelerated into this region, to prevent interaction with the
laser light before they reach the FDR. Otherwise, multiple
excitation will in most cases lead to optical pumping
into dark states and hence diminish the detection sensitivity. Moreover, for two-level systems 
a long interaction region could also lead to
changes in the ion velocity and therefore to undesired shifts in the observed resonance
frequency.\\
Acceleration of the ions leaving the hot ion source region has two important consequences for CLS: First, it
leads to a strong compression of the momentum distribution in the longitudinal
phase space \cite{Kau76}. The Doppler width $\Delta \nu_{D}$ of an atomic
transition with frequency $\nu _{0}$ is normally caused by the thermal
distribution of kinetic energies $E = m \upsilon^{2}/2$ of the atoms. For an
ion beam the width of the energy distribution $\delta E$ is determined by
properties of the ion source \textit{e.g.}, the potential distribution inside
the source and the ionization process.
By acceleration in an electrostatic field the energy spread
\begin{equation}
\Delta E=m~\upsilon ~\Delta \upsilon
\end{equation}
remains constant. Consequently, if the velocity $ \upsilon $ increases, the
velocity spread $\delta \upsilon $ has to decrease accordingly. By this effect,
at energies of a few 10~keV, the Doppler width in the beam direction
\begin{equation}
\Delta \nu_{D}= \nu _{0}~\Delta \upsilon/c
\end{equation}%
is expected to be reduced to the order of the natural linewidth of a typical allowed
dipole transition. With $\upsilon = \sqrt{2eU/m}$ given by the acceleration
voltage $U$ one obtains
\begin{equation}
\Delta \nu_{D}= \nu _{0} \frac{\Delta E}{\sqrt{2eUmc^2}}\,.
\end{equation}%
The ion source conditions also influence the total energy of the
accelerated beam. This has important consequences for high-voltage measurements
and the accuracy in CLS: The potential applied externally to the source is not
necessarily identical with the potential at which ions are created. Three ion
sources are typically used at ISOLDE: surface ion sources, the resonant ionization laser
ion source (RILIS) and plasma ion sources. While for the former ones the
uncertainty of the start potential of the ions is less than 3~V, in the
plasma ion sources it can be an order of magnitude larger. Hence, high-voltage
measurements more accurate than about 20~ppm are inappropriate and will not
help to obtain higher accuracy. All measurements reported here were performed
with ions from RILIS.\\
The second consequence of acceleration is a large Doppler shift of the atomic transition
frequency according to
\begin{equation}
\nu _{\mathrm{c}}=\nu _{0}\gamma \left( 1 \pm \beta \right)
\label{nuc}
\end{equation}
in the case of collinear excitation and anticollinear excitation, respectively. Here, $\gamma = \sqrt{ 1-\beta
^{2}}$ is the relativistic time dilatation factor and $\beta
=\upsilon /c$ is the ion velocity in units of the speed of light. The
standard procedure in CLS is as follows: The laser frequency is fixed at $%
\nu _{L}$ close to the Doppler-shifted resonance frequency $ \nu _{c}$ and the velocity
of the ions is changed until the varying Doppler shift of the laser
frequency brings the atomic transition into resonance with the laser light.
The atomic resonances of isotopes with masses $m_{1}$ and $m_{2}$ occur at
different voltages $U^{(1)}$ and $U^{(2)}$\ fulfilling the condition%
\begin{flalign} 
\nu _{L}& =\nu _{0}^{(1)}\frac{m_{1}c^{2}+eU^{(1)}\pm \sqrt{%
eU^{(1)}(2m_{1}c^{2}+eU^{(1)})}}{m_{1}c^{2}} \\
& =\nu _{0}^{(2)}\frac{m_{2}c^{2}+eU^{(2)}\pm \sqrt{%
eU^{(2)}(2m_{2}c^{2}+eU^{(2)})}}{m_{2}c^{2}}  \notag
\end{flalign}%
with $\nu _{0}^{(2)}=\nu _{0}^{(1)}+\delta \nu _{\mathrm{IS}}^{1,2}$, where
$\delta \nu_{\mathrm{IS}}^{1,2}$ denotes the isotope shift. 
One should remember that $eU^{(i)}$ refers to the beam energy and not
necessarily to the applied ion source potential. We will first consider the
case that the complete acceleration voltage is measured with a miscalibrated voltage
divider. Using the approximation $\nu _{0}^{(1)}\approx \nu _{0}^{(2)}$,
because $\delta \nu $ is usually small compared to $\nu _{0}$, we obtain
\begin{equation}
U^{(2)}=\frac{m_{2}}{m_{1}}U^{(1)}.
\end{equation}
A common calibration factor in the voltage measurement will cancel out on both
sides. Hence, only a small error occurs due to the
slightly different resonance frequencies which even in the case of
beryllium with a large isotope shift and a small mass is less than 1~MHz for a $%
10^{-4}$ voltage deviation.
However, it is standard in CLS - because mass separators are used - that the total acceleration voltage is
composed of the source potential which is used to generate the fast ion
beam, and a smaller variable potential at the fluorescence detection region used for Doppler tuning. The resulting
influence of voltage errors is much larger in this case. We first solve the relation $\nu _{0}^{(2)}=\nu _{0}^{(1)}+\delta \nu _{\mathrm{IS}}^{1,2}$ to
extract a fully relativistic formula for the isotope shift and obtain%
\begin{flalign}
\delta \nu_{\mathrm{IS}}^{1,2} &=& \frac{\nu _{0}}{\kappa} \left\{ m_{2} \left( eU^{(1)} + \sqrt{eU^{(1)}(2m_{1}c^{2}+eU^{(1)})} \right) \right.  \notag\\
&&-\left. m_{1} \left( eU^{(2)} + \sqrt{eU^{(1)}(2m_{1}c^{2}+eU^{(1)})} \right) \right\} 
\end{flalign}
where $\kappa = {m_{1}\left[ m_{2}c^{2}+eU^{(2)}+\sqrt{eU^{(2)}(2m_{2}c^{2}+eU^{(2)})}\right]}. $
The effect of a calibration error in the isotope shift measurement can now
easily be calculated by replacing $U^{(1)}$ and $U^{(2)}$, with voltages
\begin{equation}
U^{(i)\prime}  = U^{(i)}_{\mathrm{Source}}(1+k_{\mathrm{Source}})+U^{(i)}_{\mathrm{FDR}}(1+k_{\mathrm{FDR}})\,,
\end{equation}
where $i=1,2$. The unprimed values are the real voltages and the primed values
are the measured voltages, with miscalibration factors $k_{\mathrm{Source} }$
for the source and $k_{\mathrm{FDR}}$ for the fluorescence detection region. In
Table \ref{TABrequirements}, the results of these calculations are summarized
under the following assumptions which are typical for measurements at ISOLDE
but also for CLS at other facilities: The ISOLDE ion source potential is assumed to be
$U_{\mathrm{Source}}=50$~kV and the laser frequency is chosen in such a way
that the resonances of two isotopes with mass numbers $A$ and $A+2$ occur
symmetrically at post-acceleration voltages $+U_{\mathrm{FDR}}$ and
$-U_{\mathrm{FDR}}$. At COLLAPS the post-acceleration voltage $U_{\mathrm{FDR}}$ can
be up to $\pm10$~kV. For the measurement this voltage range depends on the hyperfine
structure to be scanned. For the ISOLDE miscalibration factor we assume
$k_{\mathrm{Source}} = 0.0001$ which lies within the specified accuracy of
10$^{-4}$. We further assume that $U_{\mathrm{FDR}}$ is measured exactly,
hence, $k_{ \mathrm{FDR}} = 0$. The resonances of both isotopes therefore
appear at measured voltages $U^{(i)\prime} = 50005~\mathrm{V} \pm
U_{\mathrm{FDR}}$. With these conditions the approximate error formula
\cite{Mue83} for the isotope shift becomes
\begin{equation}
\Delta \delta \nu_{\mathrm{IS}}^{1,\,2} = \nu_0 \sqrt{\frac{eU}{2mc^2}}
\left(\frac{U_{\mathrm{FDR}}^{(1)} - U_{\mathrm{FDR}}^{(2)}}
{2U_{\mathrm{Source}}} + \frac{(m_2-m_1)}{m_1+m_2}\right)\ k_{\mathrm{Source}}
\end{equation}
We have calculated the effect for three elements that have been recently
investigated at ISOLDE. The most crucial case is beryllium. Here, an accuracy
of about 1~MHz is required to extract the nuclear charge radius of the neutron halo
isotope $^{11}$Be \cite{Noe09} with sufficient accuracy and it is obvious that
this accuracy cannot be reached in the standard approach. For magnesium, the
voltage accuracy of 10$^{-4}$ barely meets the requirements, while in the case
of copper the standard voltage divider is sufficient. If charge radii for two or more
isotopes are available from other techniques, for example muonic atom spectroscopy, the requirements are slightly less severe. The systematic error caused by the voltage miscalibration behaves largely as the mass shift. Hence, the King plot procedure used in such cases, delivers a mass shift constant that will result in a correct field shift evaluation. However, for the lightest nuclei even this approach does not provide sufficient accuracy.
\begin{table*}
\caption{
\label{TABrequirements}
Calculations of the influence of high-voltage calibrations on isotope shift measurements for light and medium mass nuclei. Ion masses are used as integer numbers, the transition frequencies refer to the $2s_{1/2}~\rightarrow 2p_{1/2}$, the $3s_{1/2}~\rightarrow 3p_{3/2}$, the $3s_{1/2}~\rightarrow 3p_{1/2}$ transition in Be$^{+}$, Mg$^{+}$ and Cu$^{+}$, respectively. $\delta \nu _{\mathrm{IS}%
}^{A,A+2}$ is the complete isotope shift between the two isotopes with mass numbers A$_{1}$ and A$_{2}$. The differential isotope shift $ \partial\nu
/\partial U(A)$ was calculated according to Eq.(~\ref{eqdiffis}) and $\Delta \delta \nu _{\mathrm{IS}}^{A,A+2}$ gives the artificial isotope shift introduced by a 10~$^{-4}$ measurement error on the acceleration voltage. The last column indicates the typically required accuracy on the isotope shift measurement to extract charge radius information. For more details see text.
}
\begin{center}
\begin{tabular}{@{}llllll} \hline
  Isotope pair & $\nu _{0}(A)$ & $\delta \nu _{\mathrm{IS}}^{A,A+2}$ & $\partial \nu /\partial U(A)$ & $\Delta \delta \nu _{\mathrm{IS}}^{A,A+2}$ & Required accuracy \\
   & (THz) & (MHZ) & (MHz/V) & (MHz) & (MHz) \\ \hline \hline
   $^{9-11}$Be & 957.22 & 14240.0 & 31.2 & 29.2 & 1 \\ \hline 
   $^{24-26}$Mg & 1072.08 & 3220 & 22.5 & 8.5 & 3  \\ \hline
   $^{63-65}$Cu & 923.01 & 2000 & 11.9 & 1.7 & 5  \\ \hline
   \hline
   \end{tabular}
   \end{center}
\end{table*}

In summary, CLS of light elements is difficult because the ions have a large
differential Doppler shift
\begin{eqnarray} 
\frac{ \partial \nu }{\partial U} & = &  \frac{\nu_{0}}{mc^{2}} \left( e+\frac{e\left( mc^{2}+eU\right) }{\sqrt{eU \left(2mc^{2}+eU \right) }}\right) \nonumber \\
& & \approx \frac{e \cdot \nu_{0}}{\sqrt{2 e U m c^{2}}}
\label{eqdiffis}
\end{eqnarray}
and the observed resonance positions are therefore extremely sensitive to
the applied voltage and thus to voltage fluctuations. Moreover, the mass shift in
these isotopes is huge compared to the tiny nuclear volume effects.
Nevertheless, a few techniques were developed to overcome these problems and
allowed CLS of a few, particularly interesting cases. CLS of short-lived neon
isotopes including the two-proton halo candidate $^{17}$Ne was performed using a
special feature of the neon atomic spectrum: Two fine-structure transitions
in neon starting from a common metastable level, which was populated in
charge exchange collisions, have a splitting that fits exactly to the
opposite Doppler shifts in collinear and anticollinear excitation at a beam
energy of about 61.8~keV if the laser frequency is set to the average
transition frequency in the rest frame. This was used in the neon
measurements at COLLAPS to calibrate the beam energies and to obtain
reliable isotope shift values for the neon isotopes \cite{Get00,Gei08}. In this
experiment, good agreement between the applied high voltage at ISOLDE and
the beam energy was found. On Li$^{+}$ ions Riis and coworkers \cite{Rii94}
used collinear and anticollinear saturation spectroscopy, which is not sensitive to an exact match of the ion velocity
with the optical resonance. Such a type of spectroscopy is also used for Ives-Stilwell tests of
Special Relativity at storage rings \cite{Rei07,Nov09}. Here, the rest frame
frequency is obtained from the observed Lamb dips using the product formula
of Eq.~(\ref{nuc}):
\begin{equation}
\nu _{0}^{2}=\nu _{\mathrm{c}}\cdot \nu _{\mathrm{a}},
\label{eq:nurest}
\end{equation}%
with $\nu _{\mathrm{a}}$ and $\nu _{\mathrm{c}}$\ being the laser
frequencies at the resonance position in anticollinear and collinear
geometry, respectively. This relation is independent of all applied voltages
but requires knowledge of the absolute laser frequencies typically on the $%
10^{-9}-10^{-10}$ scale. In the lithium experiment \cite{Rii94}, this was performed using
iodine reference lines and the isotope shift of the two stable isotopes $%
^{6,7}$Li was extracted with good accuracy.\\
To avoid uncertainties of
the voltage measurement and to determine the isotope shift between $%
^{7,9,10,11}$Be, as well as the absolute excitation frequencies $\nu _{0}$
for the $2s_{1/2}\rightarrow 2p_{1/2,3/2}$ transitions with an accuracy 
better than 2~\textrm{MHz}, a similar approach was used for the first
time in an on-line experiment. A detailed discussion of the laser system and experimental setup is given in
\cite{Noe09, Zak10}. Summarized, as shown in Fig.~\ref{collsetup}, two dye lasers were collinearly and
anticollinearly overlapped with a beryllium ion beam. For precise frequency
determination and stabilization one laser was locked to a
frequency comb, the second one to an iodine transition via Doppler-free
frequency modulation saturation spectroscopy in an iodine cell. From two second-harmonic generation
devices, UV light of 312~nm and 314~nm was coupled through quartz windows into
the collinear and anticollinear directions of the ion beam, respectively.
The observed resonance frequencies  $\nu _{\mathrm{a}}$ and $\nu _{\mathrm{c}}$ allowed
us to determine the absolute transition frequency $\nu _{\mathrm{0}}$ in the rest frame of
the ion according to Eq.~(\ref{eq:nurest}). The isotope shift was then extracted as the difference of the absolute transition
frequencies between different isotopes. Then the inversed process was used and from the now known value of $\nu _{0}$ for $%
^{9}$Be and the collinear and anticollinear signal
the beam energy was determined with an accuracy at the level of $10^{-5}$ using Eq.~(\ref{nuc}).\\ 
\newline
The relative difference between the beam energy and the
ISOLDE high-voltage readings were found to be $-2.0(2) \cdot 10^{-4}$ for the power supplies HT1 and $6.5(4) \cdot 10^{-4}$ for HT2, corresponding to absolute values of $-12(1.3)(3.0)$~V and $39.0(2.4)(3.0)$~V at the maximum operation voltage of 60~kV. Here the first uncertainty denotes the pure fitting uncertainty and the second one the systematic uncertainty. This systematic uncertainty accounts for the limited knowledge about the potential at which ionization takes place in the source. Both devices, including the high-voltage dividers, were installed at ISOLDE in 1981 and are in operation since then. More details will be given in Section \ref{sec:isolde}.  \\
Since we found a clear discrepancy between the measured ISOLDE
ion source voltages and the beam energy, we attributed this to a problem in
the high voltage readout. Therefore, a recalibration of the
ISOLDE high-voltage divider was arranged for future CLS experiments using an independent high-precision voltage divider.\\
\section{The high-voltage installation at the ISOLDE facility}
\label{sec:isolde} 
\begin{figure}
\centering
\includegraphics[width=0.85\linewidth]{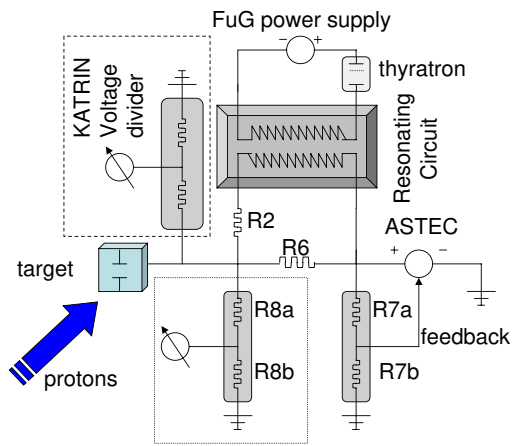}
\caption{Simplified schematic view of the high voltage circuitry at the two ISOLDE HV units including a resonant circuit used to modulate the high voltage applied to the target. The resistors R7a/R7b represent the ROSS voltage divider for internal stabilization whereas R8a/R8b (framed) represent the voltage readout $U_{\mathrm{ISOLDE}}$.}
\label{lb:rescircuit}
\end{figure}

The on-line mass separator ISOLDE at CERN produces low-energy radioactive ion beams \cite{Kug02}
making use of the so called ISOL\footnote{isotope separation on-line} technique: A stack of four synchrotrons, the Proton Synchrotron Booster (PSB), produces a pulsed
1.4~GeV proton beam. These high-energy protons are delivered to one of the two target
stations, inducing nuclear reactions in a thick target to produce radioactive isotopes.
The reaction products are ionized, $e.g.$ via resonant laser ionization at about \mbox{2000 °C}
\cite{Klu85,Mis92}. The targets and ion sources are mounted on two platforms of positive electrical
potential up to 60~kV. The ions are then extracted and accelerated to ground potential for mass
separation. For that purpose either the General Purpose Separator GPS or the
High Resolution Separator HRS magnetic sector separator can be used.
The mass separated ion beam is  guided to the experiments.\\
The target and ion source platform is connected to either of two units HT1 or HT2 comprising
a stabilized HV power supply and a voltage divider for the readout of the applied ion source potential.
The high voltage circuitry of each unit is shown in Fig.~\ref{lb:rescircuit}.
Up to 2008 the acceleration voltage was generated by two actively regulated ASTEC 60~kV precision
power supplies. The accuracy of both was specified to $10^{-4}$. The internal voltage
stabilization feedback loop is based on a ROSS voltage divider 75-10-BDL, represented by resistors R7a/R7b in Fig.~\ref{lb:rescircuit},
using a scale factor 1000:1. In order
to stabilize the devices in temperature to $\pm 0.1$~K the power supplies, the resonant circuit and the
ROSS voltage divider were immersed in an oil bath.\\
A similar ROSS voltage divider at a scale factor of 10000:1, represented by resistor R8a/R8b 
and a high-precision HP 3458A digital multimeter (6~$1/2$ digit resolution, which corresponds to an accuracy of $10^{-5}$)
are used to read out the stabilized target voltage to the ISOLDE control system via GPIB. These dividers were originally specified with an accuracy of $10^{-4}$ and a stability of  $5 \cdot 10^{-5}$ per year.\\
In 2009 the former ASTEC power supply of HT1 was replaced by a self-regulated
Heinzinger power supply of the PNChp3p series. In this commercial device
the applied voltage is regulated internally and specified in stability to $1 \cdot 10^{-5}$ over a time
period of 8 hours under constant conditions and in load regulation, and for a load step from zero to
the nominal voltage, to $5 \cdot 10^{-4}$. For readout the ISOLDE control system
uses again the ROSS voltage divider and HP 3458A multimeter, {\it{i.e.}}, resistors R8a/R8b in Fig.~\ref{lb:rescircuit}.\newline
During proton beam impact, the ionization of the air volume around the target gives rise to a significant leakage current which results in loss of charge on the effective target capacitance. This voltage breakdown could lead to damage of the HT supply and furthermore can significantly slow down the regulation loop, hence the reason of blanking the high voltage by a resonant circuit \cite{Fia92}. The principle task of this resonance circuit is to fully
discharge the target capacitance prior to beam impact. The schematic is included in the upper right part of Fig.~\ref{lb:rescircuit}: 
The first pole of the pulse transformer is connected to a commercial power supply (FuG Elektronik, 0-15~kV),
whose output voltage is a function of the ASTEC's output voltage, controlled by a thyratron, which is synchronously triggered with the PSB 
ejection kickers. The current supplied by the resonant
circuit discharges through the capacitance of the target. Thus, a high-voltage pulse of up to 60~kV
with opposite polarity modulates the target voltage (within 30~$\mu$s) to zero, just before the
protons collide with the target. A small resistor R2 (=250~$\Omega$) reduces the stabilization time
to less than 10~ms after the pulse for $\pm$~1~V from the nominal voltage. At this point a voltage drop
of R2$\cdot$I is at most 500~mV. The resistor R6 (=1.5~$ k\Omega$) forms a fast coupling of the ASTEC supply to the target.
Note, that the power supply itself is running in a static mode and
that only the acceleration gap receives the modulated voltage. The figure shows also the connection to the high-precision Karlsruher Tritium Neutrinoexperiment (KATRIN) dividers
used for calibration.\\
\begin{figure}[ht]
\centering
\includegraphics[width=\linewidth]{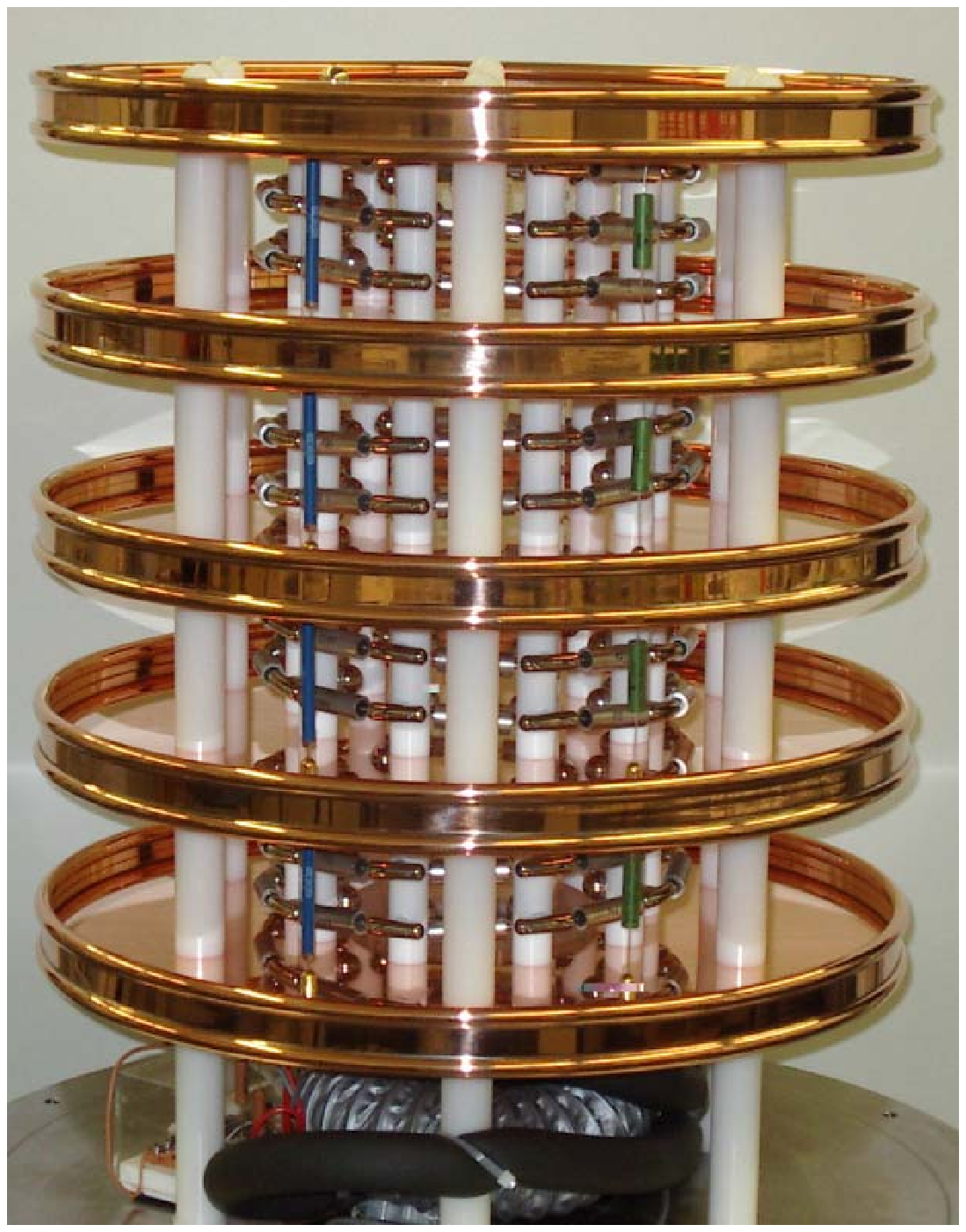}
\caption{The interior of the high-voltage divider K1 from the Karlsruher Tritium Neutrinoexperiment (KATRIN) experiment: 106 precision resistors with a total resistance of 184~M$\Omega$ are arranged in a helix structure around five copper ring electrodes and connects each pair of copper electrodes to provide a linear potential distribution over the primary high-precision resistor chain. Each pair of copper electrodes is connected by a high voltage resistor (blue) and capacitor (green) to control the electrical potential over the primary resistor chain and to protect it against overloads.}
\label{HVdivSetup}
\end{figure}
\section{The High-Precision Voltage Divider}
\label{sec:K1} Although small voltage differences can be measured to the
highest precision, $e.g.$ $10^{-12}$ applying the Josephson effect \cite{Ham00,Jos62}, measuring high-voltages at precision levels below $10^{-4}$ becomes increasingly difficult. To measure
voltages larger than 10~kV with high accuracy, voltage dividers are required
to downsize the voltages to a range accessible for a precise digital multimeter.
A high-precision voltage divider, further denoted K1 \cite{Tue07},
was developed for the Karlsruher Tritium Neutrinoexperiment KATRIN \cite{Bec09}, which is able to
determine voltages up to 35~kV in the sub-ppm regime. The KATRIN experiment measures the mass of the electron neutrino directly with highest sensitivity in the sub-eV regime using a so-called MAC-E-Filter (magnetic adiabatic collimation combined with an electrostatic filter).  Thus high luminosity and energy resolution are combined in order to determine the energy distribution of the electrons at the end-point of a tritium ß-decay. The precision of this energy measurement relyes on the ppm precision of the applied electrostatic retarding potential.\\
One year later, in 2009, a second high-precision divider K2 for the KATRIN experiment was completed to measure high voltages up
to 65~kV. Both dividers were specified at the Physikalisch Technische Bundesanstalt (PTB) to sub-ppm accuracy and stability. We report here on a voltage calibration of the two ISOLDE high
voltage units HT1 and HT2 against these high-precision voltage dividers.\\
Figure \ref{HVdivSetup} shows a picture of the K1 divider: A cylindrical stainless steel container with a diameter of 60~cm and a height of 85~cm houses the voltage divider setup. The case, which shields from electromagnetic interference from outside, is flooded with N$_{2}$ gas
at atmospheric pressure along the resistors for insulation and temperature stabilization at $25 \pm 0,15~$C. The interior consists of five copper
ring electrodes fixed by polyoxymethylene rods which are concurrently used as insulators. The heart of the voltage divider
is the primary resistor chain, which consists of 106 precision 
resistors of Bulk Metal Foil 
technology\footnote{BMF is a brand of Vishay Intertechnology Inc.:www.vishay.com} with a total
resistance of 184~M$\Omega$ \cite{Tue07}. These resistors are arranged in a helix 
structure between the copper electrodes. Up to the forth one, each
ring comprises 25 resistors. The low voltage output is formed by the remaining six resistors arranged in each case by three
resistors in parallel. Thus a scale factor of 3945:1 or 1972:1 can be provided. The copper electrodes generate a constant
electric field along the primary high-precision resistor chain. A secondary resistor chain of four high-voltage 44~M$\Omega$ resistors (Caddock MX480) connects each pair of cooper electrodes to provide a linear potential distribution. In order to protect the primary resistors against transient overloads a capacitive chain, consisting of four 2.5~nF capacitors (Vishay MKT1816), is implemented between the copper electrodes. The high voltage to be measured is fed into a sealed HV bushing located at the top electrode.\\
The first version of the voltage divider setup K1 used in 2008 and described above, was limited to voltages up to 35~kV. The development of a further voltage divider K2 was strained to exceed K1 in disciplines of long-term stability and temperature stabilization behavior. Since the electric strength has already been increased to 65~kV (it is conceivable that potentials up to 100~kV could be measured, since the electric strength of the primary resistor chain is 100~kV, but the design would have to be modified and tested), the K2 divider is suitable to cover the full range of the acceleration potential of the ions at ISOLDE. As both dividers are identical in concept, we mention only the major changes in the setup of K2:
The copper ring electrodes divide the setup into five (instead of four) sections, where each section comprises 34 precision resistors of 880~k$\Omega$. In the last, sixth level the low voltage outputs provide divider ratios of 3636:1, 1818:1 and 100:1. The first one is used for the calibrations reported here. The secondary divider chain connects the copper electrodes and comprises 5 resistors (Caddock MX480), each with 36~M$\Omega$ and a 90~k$\Omega$ resistance (Caddock MS260) in the sixth subsection for the voltage output. The tertiary capacitive chain consists of a series of 5  capacitors  of 7.5~nF (3 capicators of 2.5~nF in parallel).\\
Both voltage dividers were specified at the PTB (see Table \ref{tab:k1-specs}).
\begin{table*}
\caption{
\label{tab:k1-specs}
  Calibration parameters of the Karlsruher Tritium Neutrinoexperiment (KATRIN) dividers K1 and K2 derived at Physikalisch Technische Bundesanstalt (PTB) \cite{Tue07}. 
  A first calibration campaign at PTB demonstrated that divider K2 shows even higher
  stabilities of the scale factor with respect to time, voltage and temperature, 
  but not all investigations are finished yet, therefore the K2 
  calibration values are still preliminary. For details see text.
}
\begin{tabular}{@{}lrr} \hline
  Parameter & divider K1 & divider K2\\ \hline \hline
  Absolute scale factor $K_0$ as of Nov 2006 (K1), Dec 2009 (K2) & $ 3944.9597(14)$ & $3636.2743$ \\ \hline 
  Long-term stability of scale factor $c_t := \frac{\partial K}{\partial t} \cdot \frac{1}{K}$ &
  $ 6 \cdot 10^{-7}$~month$^{-1}$  \\ \hline
  Temperature dependence of scale factor at $T=25~^\circ$C \quad $c_T := \frac{\partial K}{\partial T} \cdot \frac{1}{K}$ &
  $1.7(7) \cdot 10^{-7}$~K$^{-1}$ & $1 \cdot 10^{-7}$~K$^{-1}$  \\ \hline
  Voltage dependence of scale factor at $U=18.6$~kV \quad $c_U := \frac{\partial K}{\partial U} \cdot \frac{1}{K}$ &
  $-2.8(2) \cdot 10^{-8}$~kV$^{-1}$ & $-1.9(2) \cdot 10^{-8}$~kV$^{-1}$ \\ \hline
  Quadratic voltage dependence $c_{UU} := \frac{\partial^2 K}{\partial U^2} \cdot \frac{1}{K} $ &
  $-7.5(4) \cdot 10^{-10}$~kV$^{-2}$\\ \hline \hline
\end{tabular}
\end{table*}
The divided voltage is measured by a FLUKE 8508A digital reference multimeter. In order to avoid drifts within this precision voltmeter, a 10~V reference potential was provided by a Fluke 732A reference voltage source, that also had been calibrated at the PTB. Before every calibration measurement, the digital voltmeter was calibrated with this 10~V reference.\\
\begin{figure*}
\centering
\includegraphics[width=\linewidth]{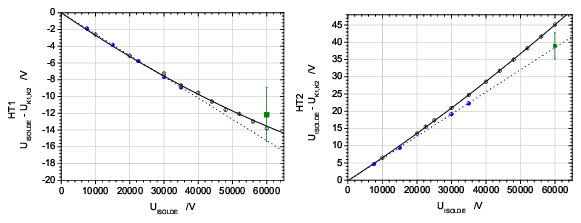}
\caption{Discrepancy between the readout of the ISOLDE high-voltage power supplies HT1 and HT2 and the voltage determined by
the KATRIN voltage dividers K1 ($\bullet$ up to 35~kV, 2008) and K2 ($\circ$ up to 60~kV, 2009). Parabolic functions are fitted to the 2009 data
and linear functions (dashed line) to the 2008 data. 
The latter was extrapolated to 65~kV. The result of a beam energy measurement at 60~kV using collinear laser spectroscopy is shown for
comparison ($\blacksquare$ at 60~kV, 2008).}
\label{calheinz}
\end{figure*}
\begin{figure*}
\centering
\includegraphics[width=\linewidth,height=0.5\linewidth ]{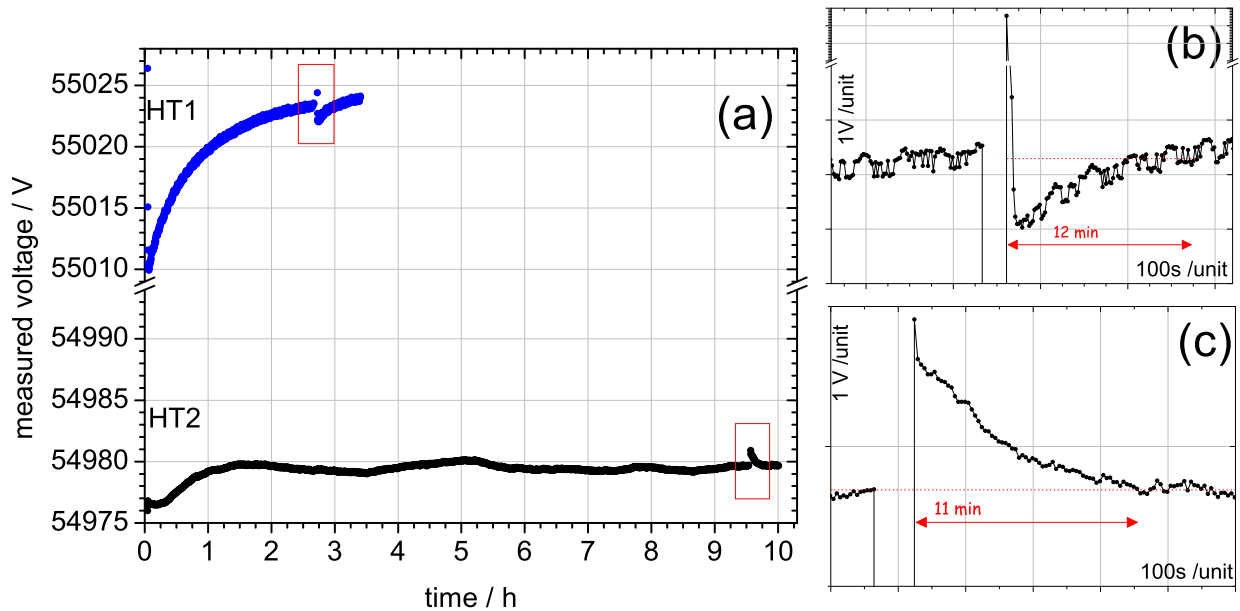}
\caption{(a) Long-term behavior of the high-voltage unit HT1 at ISOLDE equipped with the Heinzinger power supply (upper trace, top axis 3~h) and HT2 equipped with the ASTEC device (lower trace, bottom axis 10~h) for a set potential of 55~kV recorded with the KATRIN K2 divider. A short abrupt voltage drop as it occurs at ISOLDE occasionally was induced on purpose in the regions indicated by the boxes, to investigate the transient behavior after resetting the HV. This region is enlarged in the graphs (b) for HT1 and (c) for HT2.}
\label{longterm:sc}
\end{figure*}
\section{High-voltage measurements}
\label{sec:calibration}
The following section describes the calibration measurements as well as the
investigation of the transient behavior of the power supplies. Finally, the
importance of the calibration is demonstrated with a result of collinear laser spectroscopy on stable
magnesium isotopes. 
\subsection{Calibration measurements}
\label{sec:calimess} 
For calibration of the high-voltage units HT1 and HT2 the voltage
readout of ISOLDE was compared to the simultaneously measured voltage with the
precision voltage divider K1 in 2008 and K2 in 2009. Both power supplies were switched on more than three hours before the measurement started, to make sure that thermal equilibrium was reached.
The output of the power supply was then
raised stepwise from 10~kV to 35~kV (K1) and to 60~kV (K2)
and decreased again. After setting a new voltage, the system was
given a period of 7 minutes to thermalize, before 3 data points
were taken in a sequence of 30 seconds.\newline
The high voltage $U_{\mathrm{DIV}}$ is given by the scale factor of the divider $K$ times the low-voltage 
reading $U_{\mathrm{lv}}$ of the digital voltmeter (DVM). The latter is corrected for an offset $\delta U_{\mathrm{lv}}$ and a 
gain factor $k$ close to 1, which is
determined by the reference voltage source.
\begin{equation}
  U_{\mathrm{DIV}} = K \cdot (U_{\mathrm{lv}}-\delta U_{\mathrm{lv}}) \cdot k,
\end{equation}
resulting in a voltage uncertainty of
\begin{equation} 
  \frac{\Delta U_{\mathrm{DIV}}}{U_{\mathrm{DIV}}} = \sqrt{ \left( \frac{\Delta K}{K} \right)^2 + 
                              \left( \frac{\Delta k}{k} \right)^2 + 
                              \left( \frac{\Delta U_{lv}}{U_{lv}}\right)^2 } \approx \frac{\Delta K}{K}.
\end{equation}
This simplification is allowed
because the relative uncertainties of the gain factor $k$ of the DVM and of the voltage reading $U_{\mathrm{lv}}$ 
are smaller than $10^{-6}$.

The scale factor of the divider $K$ is derived from the factor obtained at the calibration date $K_0$ according to 
\begin{equation}
  K = K_0 \cdot (1 + c_t \cdot \delta t + c_T \cdot \delta T + c_U \cdot \delta U + c_{UU} \cdot \delta U^2)
\end{equation}
by considering its long-term drift $c_t$ over a time interval $\delta t$ after the last calibration,
its temperature and its voltage dependence with 
$\delta T$ being the temperature difference to the reference temperature of 25~$^\circ$C, 
and $\delta U$ being the voltage difference to the reference voltage 18.6~kV (see Table \ref{tab:k1-specs}).
The terms $c_T  \cdot \delta T$, $c_U \cdot \delta U$ and $c_{uu} \cdot \delta U^2$ are smaller than $10^{-6}$, 
therefore these corrections are neglected in our voltage measurements. Contrary, the long-term drift 
of divider K1 has been determined to be $c_t(\mathrm{K1}) = 6 \cdot 10^{-7}$~month$^{-1}$ (see Table \ref{tab:k1-specs}). 
Therefore the correction $c_t \cdot \delta t$ is significant ($\delta t \approx 24$~months) and is applied to our measured values. For
divider K2 no long-term drift has been found yet and since $\delta t = 1$~month no correction is applied. To be conservative we assume a total long-term drift 
correction uncertainty of $\Delta (c_t \cdot \delta t) = 3 \cdot 10^{-6}$ for both dividers K1 and K2.
The uncertainty of the high-voltage determination by the dividers K1 and K2 can be calculated as
\begin{eqnarray} 
  \frac{\Delta U_{\mathrm{DIV}}}{U_{\mathrm{DIV}}} &=&  \frac{\Delta K}{K} \nonumber \\
                     &\approx& \sqrt{ \left( \frac{\Delta K_0}{K_0} \right)^2 +( \Delta (c_t \cdot \delta t))^2 } \nonumber \\
                     &=& 4 \cdot 10^{-6} \label{eq:diverror}
\end{eqnarray}
The differences between the ISOLDE HT1 and HT2 readout and that of the KATRIN K1
and K2 dividers are plotted in Fig.~\ref{calheinz} as a function of the applied voltage. Each data point is the average value of three subsequent measurements as described above. The tiny error bars, mostly covered by the data points, show
the standard deviation of these 3 points. Note that in 2008 the calibration could only be carried out
up to 35~kV due to the limited operating range of the K1 divider, while the 2009 calibration was performed up to 60~kV directly using 
K2. Furthermore, after the measurements in 2008 the ASTEC power supply of HT1 was replaced with the conceptionally modern Heinzinger device.\\
The linear function $y = ax$ between the voltage deviation $\Delta U$ and the ISOLDE reading $U$ is obtained from a fit to the 2008 data ($\bullet$) and is given by\\ 
\begin{equation}
\Delta U = -2.540(24)(40) \cdot 10^{-4} \cdot U.\\
\label{eq:slopeH}
\end{equation}
The first uncertainty of the slope is the pure fitting error, while the second one denotes the systematic uncertainty according to Eq.~(\ref{eq:diverror}). Linear extrapolation to an operation voltage of 60~kV thus results in a deviation of $-15.24(15)(24)$~V. \\ 
This deviation can be compared to the laser-spectroscopic beam energy
measurement described in Sec.~\ref{collsec} that yielded a difference of $-12.0(13)(30)$~V at 60~kV. This data point is also included in
Fig.~\ref{calheinz} ($\blacksquare$). 
Note that laser spectroscopy determines the
ion velocity via the Doppler shift, which depends on the starting potential of the ions inside the ion source. Therefore a small
offset of about 1-3 Volts (systematic uncertainty) was expected corresponding to the potential gradient along the
electrothermically heated tube in which ionization takes place \cite{Koe03}. Within this, both the laser and the voltage divider measurements are in reasonable agreement.\\
The calibration measurements in 2009 ($\circ$) are better reproduced with a second-order polynomial function\\
\begin{eqnarray} 
\Delta U &=& -2.791(23)(40) \cdot 10^{-4} \cdot U \nonumber \\
& &  + 8.860(52) \cdot 10^{-10} \cdot U^{2}.
\label{eq:calH}
\end{eqnarray}
This yields a deviation of $-13.55(16)(24)$~V at 60~kV. Note that the systematic uncertainty of the second order term in Eq.~(\ref{eq:calH}) is negligibly small. Since the 2009 calibration agrees in the region from 0-30~kV with the one from 2008, it can be concluded that a second-order polynominal calibration curve is probably also valid for the old data and even reduces the difference between the laser-based and the high-voltage measurement.\\
The ROSS voltage dividers used with the ASTEC devices were specified in 1989 to an accuracy of $10^{-4}$ corresponding to 6~V at 60~kV. The observed discrepancy is about twice that value which  might be ascribed to aging effects over years. The fact that both measurements show the same deviation indicates that the internal regulation circuits of both power supplies are accurate.\\
The results of the investigation of the ISOLDE HT2 unit are shown in
Fig.~\ref{calheinz}. The measurements taken in 2008 with the KATRIN K1 divider are fitted with the linear function $y = ax + b$ (dashed line), yielding:\\
\begin{equation}
\Delta U = +6.432(26)(40) \cdot 10^{-4} \cdot U - 0.128(68)~V,\\
\label{eq:calA}
\end{equation}
which corresponds to an extrapolated voltage deviation of $+38.45(22)(24)$~V at the acceleration voltage of 60~kV. In this case we had to assume a linear function with a free offset to reproduce the data points. Laser spectroscopy at 60~kV yielded a deviation of $+39.0(24)(30)$~V which is in excellent agreement with the value obtained from the voltage calibration.\\
The calibration results in 2009 ($\circ$) were best reproduced by a second-order polynomial function of the form $y = a_{1}x + a_{2}x^{2} + b$, yielding\\
\begin{eqnarray} 
\Delta U &=& -0.118(39)~V + 6.51(46)(40) \cdot 10^{-4} \cdot U  \nonumber \\
&&           + 1.721(69) \cdot 10^{-9} \cdot U^{2}
\label{eq:calA2}
\end{eqnarray}
from the ISOLDE HT2 readout.\newline
Here, a deviation of $+45.13(56)(24)$~V at the maximum
operation voltage of 60~kV is observed, corresponding to a relative deviation of $8 \cdot 10^{-4}$.
As already mentioned earlier, studies with the previous
voltage divider K1 allowed measurements up to 35~kV only. Within this limit
no significant difference between a linear or parabolic fit was found. However, the K2 calibration measurements up to 60~kV are clearly incompatible with a linear function and require a second-order polynomial fit. Moreover, the 2008 and 2009 data differ considerably even in the region between 0-35~kV which indicates an appreciable instability of the ISOLDE HT2 high-voltage divider.\\
In a direct comparison applying a second-order parabolic fit to the 2008 measurement
with K1 similar to the recent measurements with the K2 divider (2009), one observes a
discrepancy of 6.7~V at the maximum voltage of 60~kV. This corresponds to a surprisingly large
change of $1.2 \cdot 10^{-4}$ over an interval of 18 months between both
calibrations. Therefore, the HT2 unit, in particular the associated ROSS divider, should not be used for collinear laser spectroscopy before further investigations of the day-to-day and long-term behavior have been performed.\\

\subsection{Voltage stability and transient behavior}
\label{sec:transient}

For all experiments depending on the ion
beam energy, for example laser spectroscopy or injection into ion traps, the stability
of the beam energy is essential. Hence the stability was
investigated in a long-term measurement, using the K2
divider in 2009. Both power supplies, HT1 and HT2, were switched on several hours in
advance and were operated at 10~kV. Then the voltage was increased in one step to
55~kV and recorded as a function of time. For both units we also
simulated a voltage dropout as it occasionally occurs at ISOLDE due to sparking or a leakage current on the target. Therefore, we switched the power supply to "standby" and after
one minute back to 55~kV.\newline
As visible on the very left of the large graph in Fig.~\ref{longterm:sc}, the Heinzinger power supply at HT1 (upper trace)
responds with a large, but short overshoot after setting a new voltage.
Within 3 hours it increases by more than 14~V and asymptotically reaches a stable voltage. A high
resolution zoom shows that the power supply regulates in steps
of about 200~mV to a temporary "stable" potential. This leads us to the
conclusion that the regulation circuit of the Heinzinger power supply itself is 
working well, but the voltage feedback measurement within the Heinzinger power supply needs
several hours to thermalize or stabilize. After the short voltage
drop, induced within the indicated box, the original potential is restored within 12 minutes. During these 12 minutes
an initial deviation of about 1~V is slowly being removed. This is clearly observable in the zoom shown in Fig.~\ref{longterm:sc}(b). \newline
The lower trace in Fig.~\ref{longterm:sc} shows a similar long-term measurement of the ASTEC voltage
stability of the HT2 unit. After turn-on, it starts smoother than the Heinzinger at HT1, with a
deviation of about 3~V from the stationary voltage which is reached within 1 hour.
Similarly to the Heinzinger, the ASTEC power supply circuit at HT2 responds to the voltage
drop with
an overshoot after reinitializing the device and takes about 11
minutes to regulate back to a maximum deviation of about 1~V from the average output voltage. The knowledge of this behavior is particularly important for laser
spectroscopy. Scans performed shortly 
after a voltage drop may systematically deviate from those taken with a well-stabilized
voltage and produce additional errors on the isotope shifts. Therefore it should be avoided to take data within about 10~min after reinitialization
of the high voltage, otherwise systematic shifts in the resonance positions may occur.

\subsection{Implications of the new calibration for laser spectroscopy measurements on magnesium isotopes}

\label{sec:inflspec} 
\begin{figure}[ht]
\centering
\includegraphics[width=\linewidth]{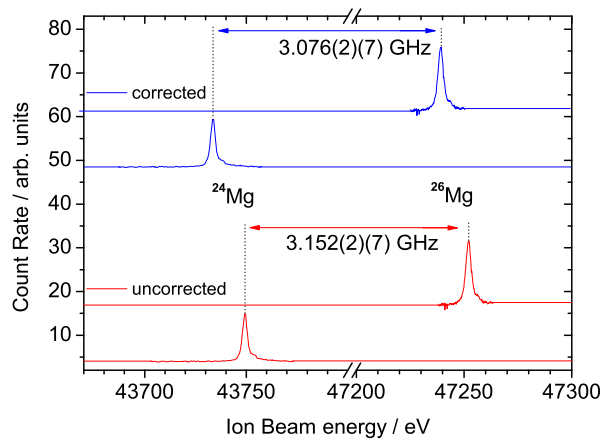}
\caption{Deviation of first isotope shift measurements on $^{24-26}$Mg
at COLLAPS from the literature value with (upper traces) and without (lower traces) including the recent 
voltage calibration.}
\label{MGISshift}
\end{figure}

The importance of the high-voltage calibration for collinear laser
spectroscopy was demonstrated in our recent isotope shift measurements on stable and short-lived Mg isotopes $^{21-32}$Mg. We measured the isotope shift in the $3s_{1/2} \rightarrow 3p_{1/2, 3/2}$ 
transitions in Mg$^{+}$ in order to investigate the rms charge radii along the isotopic chain and thus obtain information about changes in nuclear structure. The stable isotopes
$^{24}$Mg and $^{26}$Mg were chosen as reference isotopes and were repeatedly measured
during the beamtimes.\\
Resonances of $^{24}$Mg and $^{26}$Mg are presented in the lower two traces of Fig.~\ref{MGISshift}.
To obtain the isotope shift, the voltage is converted into a frequency scale and the resonance curve
is fitted with a Voigt profile. This results in an isotope shift of 3.152(2)(7)~GHz. The first uncertainty denotes the statistical and the second one the systematic uncertainty caused by the uncertainty of the start potential inside the ionizer tube of about 3~Volt.
This $^{24.26}$Mg isotope shift was recently measured with high accuracy in an ion trap \cite{Bat09},
and the reported result of $3.084905(93)$~GHz deviates from our measurement by $68$~MHz.\\
If we account for the new HV calibration, the two resonance positions are shifted in voltage as shown in the upper
two traces of Fig.~\ref{MGISshift}. The crucial point is that this shift is slightly different for the two isotopes
and the value for the isotope shift changes to $3.076(2)(7)$~GHz which is much closer to the trap measurement. The exit of the hot ionizer tube is at the ISOLDE potential \cite{Koe03} and the polarity for heating the tube is assumed to be such that the exit of the tube is more negative as it seems natural for the extraction of positive ions. Then the ion beam energy should be higher than given by the ISOLDE voltage, which shifts the IS value even closer to the trap results.\\
This demonstrates the need of a precise voltage measurement and
validates the recalibration efforts performed at ISOLDE. Particularly for the spectroscopy on light elements
a relative calibration accuracy of the order of $10^{-5}$ is required. 
\section{Summary} \label{sec:summery} 
We have calibrated the high-voltage dividers, used for measuring the acceleration voltage of ISOLDE, against the KATRIN high-precision dividers and found deviations of $2 \cdot 10^{-4}$ and $6 \cdot 10^{-4}$ for the two devices that have been in operation for more than 20 years. The
calibrations are in accordance with laser spectroscopic measurements of the ion beam energy 
using Be$^{+}$ isotopes and a frequency comb for optical frequency measurements.
Two calibrations within about 18 months have demonstrated that currently
only the unit HT1 fulfills the requirements for
collinear laser spectroscopy. For HT2 further investigations of the
day-to-day and long-term behavior are required before it can be used
again for collinear laser spectroscopy.\\
Applying the new calibration for laser spectroscopic isotope shift measurements of Mg
resolved discrepancies initially observed between collinear measurements and recent experiments 
on laser-cooled ions in a Paul trap.\\
Experiments using collinear laser spectroscopy similar to those at COLLAPS/ISOLDE are currently being prepared for example at
the TRIGA-reactor in Mainz \cite{Ket08},
which is a prototype for the LaSpec laser experiment \cite{Cam06,Rod10} at FAIR, or the BECOLA experiment \cite{Min09} at NSCL/MSU and later FRIB. At all these places the installation of a good voltage divider is foreseen in order
to ensure an accurate knowledge of the ion beam energy and to
reach the desired precision for laser spectroscopy on short-lived
radioactive isotopes.

\section{Acknowledgment}

\label{sec:Acknowledgment} This work is supported by BMBF Contract
No.06TU263I, 06UL264I, 06MZ215/TP6 and 06MZ9178I, Helmholtz Association Contract
VH-NG-148 and the EU (FP-6 EU RII3-CT-2004-506065). The KATRIN dividers are paid from money coming from our
BMBF fundings: 05CK2PD1/5, 05CK5PMA/0, 05A08PM1. Andreas Krieger
acknowledges support from the Carl Zeiss Stiftung (AZ:21-0563-2.8/197/1). We acknowledge
support from the ISOLDE technical group.

\end{document}